\documentclass[aps,pre,twocolumn,showpacs,superscriptaddress]{revtex4-1}
\usepackage{graphicx}
\usepackage{amsmath}
\usepackage{color}

\begin{document}

\title{Effect of microtubule-associated protein tau in dynamics of single-headed motor proteins KIF1A}

\author{J. Sparacino}
	\email{sparacin@famaf.unc.edu.ar}
	\affiliation{Facultad de Matem\'{a}tica, Astronom\'{i}a y F\'{i}sica, Universidad Nacional de C\'{o}rdoba  and CONICET, Medina Allende s/n, Ciudad Universitaria, 5000 C\'{o}rdoba, Argentina}

\author{M.G. Far\'ias}
	\affiliation{Instituto de Investigaci\'{o}n M\'{e}dica Mercedes y Mart\'{i}n Ferreyra. INIMEC-CONICET and FoNCyT. Friuli 2434, Barrio Parque V\'{e}lez Sarsfield,  5016 C\'{o}rdoba, Argentina.}

\author{P.W. Lamberti}
	\affiliation{Facultad de Matem\'{a}tica, Astronom\'{i}a y F\'{i}sica, Universidad Nacional de C\'{o}rdoba  and CONICET, Medina Allende s/n, Ciudad Universitaria, 5000 C\'{o}rdoba, Argentina}


\begin{abstract}
	Intracellular transport based on molecular motors and its regulation are crucial to the functioning of cells. Filamentary tracks of the cells are abundantly decorated with non-motile microtubule-associated proteins, such as tau. Motivated by experiments on kinesin-tau interactions [Dixit et al. Science \textbf{319}, 1086 (2008)] we developed a stochastic model of interacting single-headed motor proteins KIF1A that also takes into account the interactions between motor proteins and tau molecules. Our model reproduce experimental observations and predicts significant effects of tau on bound time and run length which suggest an important role of tau in regulation of kinesin-based transport.
\end{abstract}




\maketitle

\section{Introduction}

 Intracellular transport  is fundamental for cellular function, survival, and morphogenesis. 
 Kinesin superfamily proteins (also known as KIFs) are important molecular motors that directionally transport cargoes along microtubules (MTs), including membranous organelles, protein complexes and messenger ribonucleic acids (mRNAs) \cite{Howard}. 
 Disruptions or defects of MT-based transport are observed in many neurodegenerative diseases \cite{Aridor,  Toyoshima}.
 MTs are decorated with non-motile microtubule-associated proteins (MAPs) that promote MT assembly and play important roles in organizing the MT cytoskeleton  \cite{Lee, Hirokawa1994}.
 Kinesin proteins interfere with MAPs, being the latest able to inhibit active transport of cytoplasmic material. Kinesin competes with MAPs to bind to the MT surface and MAPs bound to MTs might also block the path of motor proteins.
 Tau is a mainly neuronal  MAP, enriched in the axonal compartment \cite{Papasozomenos} that has been shown to inhibit plus end-directed transport of vesicles along MTs by kinesin \cite{Ebneth}.
 Tau reduces not only the attachment frequency of kinesin to MTs but also the distance that kinesin travels along the MT in a single run \cite{Trinczek}. 
 It is also known that when single kinesin motors encounter tau patches on the MT, most of the motors detach from the MT surface \cite{Dixit}. 
 
 Some recent theoretical models of interacting molecular motors \cite{Govindan, Belitsky, Melbinger, Johann} are extensions of asymmetric simple exclusion processes (ASEP) \cite{Derrida} in which motors are represented by particles that hop along a one-dimensional lattice with hard-core exclusion. 
 A paper by Parmeggiani et al. \cite{Parme1} introduced an ASEP--like model that relaxes the restriction of the conservation of particles in the bulk, allowing the attachment or detachment of particles in any site. 
 In this paper we consider specifically the effect of MAP tau in the intracellular traffic of single-headed kinesin motor protein KIF1A \cite{Hirokawa2}. 
 A model for KIF1A motor transport has been proposed by Nishinari et al. \cite{Nishinari} that enriched Parmeggiani's model via two significant modifications. 
 Firstly, the model explicitly incorporated the Brownian ratchet mechanism for individual KIF1A motors, i.e., the mechanochemical coupling involved in their directed motion, including the adenosine triphosphate (ATP) hydrolysis that fuels the cyclic steps of the motor.
 Secondly, the model included a construction based on parameters that have direct correspondence with experimentally controllable quantities, of which the two most important are ATP and motor
concentrations. 
 In contrast to earlier models of molecular motors intracellular transport \cite{Govindan, Belitsky, Melbinger, Johann, Parme1, Nishinari}, which consider only  motor-motor interactions, our model incorporates the effect of tau on kinesin motors dynamics. 
 One remarkable exception is the paper by Chai et al. \cite{Chai1} which considers transport of molecular motors in the presence of static defects.  
 Unlike the model investigated by Chai et al., we consider  specifically MAP tau effects in kinesin dynamics via parameters that are related to experimentally observed quantities by Dixit et al \cite{Dixit}.

\section{Lattice-gas model for KIF1A dynamics in presence of tau molecules}

 We work with a one-dimensional lattice of $L$ sites that represents one MT's protofilament.
 Each site on the lattice corresponds to one motor-binding site on the protofilament. 
 The lattice parameter is taken to be $8nm$, which is the separation between adjacent binding sites of the MT. 
 Each KIF1A is represented by a particle that, when bound to one site of the filament, can hop to any of the two nearest-neighbor sites. 
 The motor protein can also attach to (detach from) the MT; this is modeled as a particle creation (annihilation) on the filament. 
 Each site of the filament has two internal states that represent the two possible bound states in which a KIF1A motor can be related to the MT. 
 In the state 1, the motor protein is strongly bound to the MT, whereas in state 2 it is weakly bound (the motor protein is tethered to the MT by an electrostatic attraction that prevents it from diffusing away from the filament) \cite{Hirokawa2}. 
 Therefore, there are three possible states for each site: 0 if the site is empty, 1 if the site is occupied with a motor in state 1, and 2 if the site is occupied with a motor in state 2. 
 We consider a distribution of tau molecules decorating the filament. 
 Each binding site of the filament can either have one tau molecule or no molecule at all. 
 To characterize the amount of tau present in the MT, we will refer to the concentration of tau molecules, denoted by $C_{tau}$, defined as the number of tau molecules per site in the filament.
 The tau concentration therefore ranges between 0 and 1, with these extreme cases representing no tau molecules attached to any site of the MT and a tau molecule attached to every site of the filament, respectively. 

 The dynamical evolution of the system is described with transition rates that reflect the stochastic nature of the movement of the motor protein and of its interaction with tau molecules. 
 We run Monte Carlo simulations that update the state of all sites of the filament following the random sequential method. 
 In one Monte Carlo Step (MCS) we give to all sites, in a random order, the opportunity to change their state of occupation.
 The  change of state of a given site is regulated by transition rates which depend on the state of occupation of the site, the state of occupation of its nearest-neighbor sites, and also depend on whether or not there are tau molecules in the sites under consideration.

\begin{figure}[htb]
	\begin{center}
		\includegraphics[width=8.6cm]{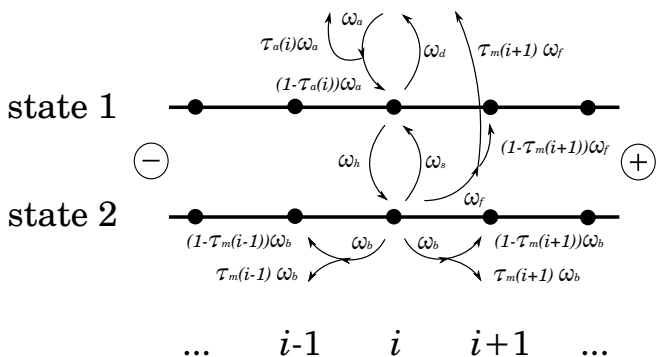}
		\caption{Scheme of the model showing the updating possibilities and the associated transition rates for a bulk site. The effect of tau, preventing a motor attachment or forcing a moving motor to detach is graphically depicted as branching arrows. 
		\label{fig:esquema} }
	\end{center}
\end{figure}

 Figure \ref{fig:esquema} shows a scheme with the updating possibilities for a site in the bulk (different rules apply for the two ends of the lattice and they are specified below). 
 Let us consider the $ith$ site: if the site is empty, the only possible state change is that a motor binds to it in state 1, with a rate $\omega_a (1-\tau_a(i))$. 
 $\tau_a(i)$ represents the probability that a tau molecule prevents the attachment of a KIF1A motor protein to site $i$. 
 It is equal to 0 if there is no tau molecule in the site, and equal to $p_a$ if there is a tau molecule in the site. 
 If the site $i$ is occupied with a motor in state 1, two things can happen. 
 Firstly, the motor protein can detach from the site, with a rate $\omega_d$, and secondly, the motor protein can hydrolyze the ATP molecule and consequently change to state 2, with a rate $\omega_h$. 
 If the site is occupied with a motor in state 2, several possibilities may arise. 
 The motor protein can diffuse to one of its nearest-neighbor sites with a rate $\omega_b (1-\tau_m(i-1))$ towards the minus end, and $\omega_b (1-\tau_m(i+1))$ towards the plus end (staying in state 2); it can release adenosine diphosphate (ADP) and move forward with a rate $\omega_f (1-\tau_m(i))$ (binding to the next site in state 1), or, finally, it can release ADP and stay in the same site with a rate $\omega_s$ (changing to state 1). 
 $\tau_m(i)$ represents the probability that a tau molecule in site $i$ forces a moving motor protein to detach.
 It is equal to 0 if there is no tau molecule in the site, and equal to $p_m$ if there is a tau molecule in the site. 
 In this updating process a change of site is possible only if the target site is empty.

 For the dynamics at the ends, we take $\alpha$ and $\delta$ instead of $\omega_a$ as the attachment rates for the minus and plus ends, respectively. In the same way, $\gamma_1$ and $\beta_1$ (instead of $\omega_d$) are the detachment rates, and $\gamma_2$ and $\beta_2$ (instead of $\omega_b$) are the exit rates of the motors due to Brownian motion for the minus and plus ends, respectively.

 Rate $\omega$  means that in an infinitesimal time interval $dt$, the probability of the event occurring is $\omega dt$.

\subsection{Mean-field equations}	

	Let us denote by $r_i$ and $q_i$  the probabilities of finding a KIF1A motor in states 1 and 2, respectively, at the lattice site $i$ at time $t$. The master equations for the dynamics of motors in a bulk site are given by:
	\begin{align} 
		\frac{dr_i}{dt} =& \: \left(1-\tau_a(i) \right) \omega_a  \left( 1-r_i-q_i \right) - (\omega_d+\omega_h)  r_i + \omega_s q_i \nonumber \\ &+ \left(1-\tau_m(i) \right)  \omega_f  q_{i-1}  \left( 1-r_i-q_i \right) \label{eq:EcMaestra1}, \\
		\frac{dq_i}{dt} =& \: \omega_h r_i - \omega_s q_i - \omega_f q_i  \left( 1-r_{i+1}-q_{i+1} \right) \nonumber \\ &- \omega_b q_i \left(2 - r_{i-1} - q_{i-1} - r_{i+1} - q_{i+1} \right) \nonumber \\ &+ \left(1-\tau_m(i) \right)  \omega_b \left(  q_{i-1} + q_{i+1} \right) \left( 1-r_i-q_i \right) \label{eq:EcMaestra2} ,
	\end{align}
where 
\begin{equation} \label{eq:tau_lig}
	\tau_a(i) =
	\begin{cases}
		p_a		& \text{if there is a tau molecule on site i} \\
		0		& \text{if there is not a tau molecule on site i} 
	\end{cases} 
	, 
\end{equation}
and

\begin{equation} \label{eq:tau_mov}
	\tau_m(i) =
	\begin{cases}
		p_m		& \text{if there is a tau molecule on site i} \\
		0		& \text{if there is not a tau molecule on site i} 
	\end{cases} . 
\end{equation}

 The corresponding equations for the minus and plus ends can be written in a similar way, with their appropriate constants as described previously.

\subsection{High-density limit}

 We can obtain an approximate solution in the high $\omega_a$ limit, which corresponds to a high motor concentration regime with a jammed region that covers the whole MT \cite{Sparacino}. 
 Assuming periodic boundary conditions, the solutions $(r_i,q_i)=(r,q)$ of the mean-field equations in the steady state, with the additionally assumption that $\tau_a(i) = \tau_a \: C_{tau} $ and $\tau_m(i) = \tau_m \: C_{tau} $, are found to be 
\begin{equation} \label{eq:r_limite}
	r = \frac{1 +  \left( K_s (1-\Omega_f)-1 \right) q }{1 + K_d + K_h (1-\Omega_f)} , 
\end{equation}
\begin{equation} \label{eq:q_limite}
	q = \frac{-A - \sqrt{A^2 - 4 \Omega_h \left( K_d + (K_h+K_s)(1-\Omega_f)  \right) }}{2 \left( K_d + (K_h+K_s)(1-\Omega_f)  \right)},
\end{equation}
where   $A = K_h (\Omega_f -1) - K_d (\Omega_s + 1) - (\Omega_s + \Omega_h) $,\\ $K_s =\frac{\omega_s}{(1-\tau_a) \omega_a}$, $K_h =\frac{\omega_h}{(1-\tau_a) \omega_a}$, $K_d =\frac{\omega_s}{(1-\tau_a) \omega_a}$,\\ $\Omega_s =\frac{\omega_s}{\omega_f+2 \omega_b \tau_m}$, $\Omega_h =\frac{\omega_h}{\omega_f+2 \omega_b \tau_m}$, and $\Omega_f =\frac{(1-\tau_m) \omega_f}{\omega_f+2 \omega_b \tau_m}$.

\begin{figure*}[htb]
	\begin{center}
		\includegraphics[width=17.2cm]{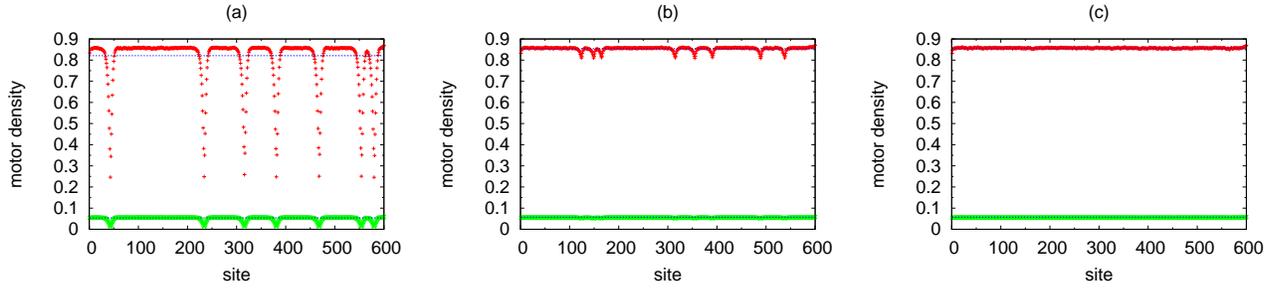}
		\caption{(Color online) Stationary density profiles for $p_a=0.67$, $p_m=0.5$ (left), $p_a=0.067$, $p_m=0.05$ (center) , and $p_a=0.0067$, $p_m=0.005$ (right). Model parameters: $L=600$, $C_{tau}=0.01$, $\omega_a=\alpha=0.001[ms^{-1}]$, $\omega_d=\beta_1=\beta_2=0.0001[ms^{-1}]$,  $\omega_s=0.145[ms^{-1}]$, $\omega_f=0.055[ms^{-1}]$,  $\omega_b=0.6[ms^{-1}]$,  and $\gamma_1=\gamma_2=\delta=0$. The red upper (green lower) points correspond to numerical results for the state 1 (state 2) density. The dashed lines are the analytical approximate predictions for periodic systems with the same parameters.  Note that the chosen $\omega_a$ value allows us to realize the high density limit \cite{Sparacino}. \label{fig:densidades} }
	\end{center}
\end{figure*}

 A system with periodic boundary conditions is not realistic. However it provides a good approximation in the high motor concentration limit with open boundary conditions.   
 Figure \ref{fig:densidades} shows a comparison between these analytical expressions and numerical simulations for three cases corresponding to different magnitudes of the tau-kinesin interactions.  

 The limit with no tau molecules present in the cytosol can be obtained taking $\tau_a=0$ and $\tau_m=0$. In that case, expressions \eqref{eq:r_limite} and \eqref{eq:q_limite} reduce to the ones found by Nishinari et al. in the model for KIF1A dynamics without tau molecules \cite{Nishinari}.

\section{Bound times and run lengths}

 We work with a MCS representing $0.71ms$ of real time. 
 Every simulation has a fixed concentration of tau molecules and starts with no motor attached to the filament. 
 In order to construct the distribution of tau molecules with the desired concentration, we give to each site of the filament the chance of having a tau molecule with probability $C_{tau}$. 
 Dixit et al. found that tau patches were stable over the time course of several minutes while the average duration of motor protein runs are in the order of seconds or tens of seconds \cite{Dixit}.
 Therefore, the tau distribution is held constant during the complete course of the simulation.
 Our model captures the essence of the experimental results on tau-kinesin interactions found by Dixit et al. \cite{Dixit}, that can be summarized as: i) the frequency of kinesin binding to MT depends on local tau concentration; ii) the average binding frequency of kinesin in presence of tau in the cytosol reduces to approximately one third of the value without tau; iii) approximately half of the times that a kinesin motor protein encounters a tau molecule, it detaches from the MT, the remaining half times it passes or pauses. Therefore, the probability that a tau molecule prevents a kinesin motor protein binding to the MT is taken to be $p_a=0.67$, and the probability that a tau molecule forces a moving kinesin motor protein to detach from the MT is chosen to be $p_m = 0.5$. All parameters that have no relation with tau were chosen following Ref. \cite{Nishinari}.

\begin{figure}[htb]
	\begin{center}
		\includegraphics[width=8.6cm]{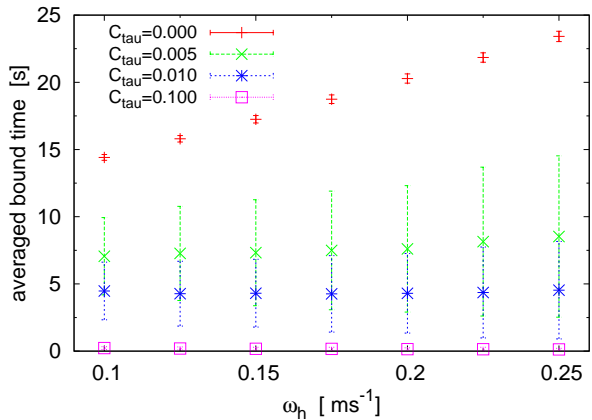}
		\caption{ Averaged time between attachment and detachment of motor proteins on the filament as a function of the hydrolysis rate. 
		\label{fig:t} }
	\end{center}
\end{figure}

\begin{figure}[htb]
	\begin{center}
		\includegraphics[width=8.6cm]{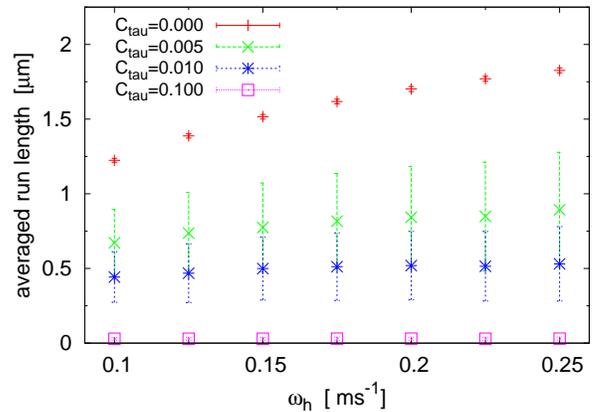}
		\caption{ Average length spanned by KIF1A motors on the filament in single runs as a function of the hydrolysis rate. 
		\label{fig:l} }
	\end{center}
\end{figure}

 For the cases shown in Figs. \ref{fig:t} and \ref{fig:l} the model parameters were taken to be $L=600$,    $p_a=0.67$, $p_m=0.5$, $\omega_a=\alpha=0.00001[ms^{-1}]$, $\omega_d=\beta_1=\beta_2=0.0001[ms^{-1}]$,  $\omega_s=0.145[ms^{-1}]$, $\omega_f=0.055[ms^{-1}]$,  $\omega_b=0.6[ms^{-1}]$,  and $\gamma_1=\gamma_2=\delta=0$.
 The observables shown were obtained averaging the results of 1000 simulations with the same parameters except for the random number generator seed. 
 The error associated to the observables correspond to the standard deviation. 
 Each one of the simulation generates a new tau distribution (with the same concentration) and evolves for 2x$10^6$ MCSs after a thermalization of 2x$10^5$ MCSs which guarantees that the system reaches the stationary state.
 Figure \ref{fig:t} shows the behavior of the bound time as a function of $\omega_h$. In each simulation the bound time is calculated by averaging the intervals between attachment and detachment of each KIF1A.
 It is worth remembering that $\omega_h$ is a monotonous increasing function of ATP concentration ($\omega_h=0.25 {ms}^{-1}$ corresponds to saturating ATP concentration) \cite{Nishinari}.
 For $C_{tau}=0$ there are no tau molecules present in the cytosol (in this case, the values of $p_a$ and $p_m$ are irrelevant) and the bound time grows with $\omega_h$ from near $15 s$ for low ATP concentration to almost $25 s$ for saturating ATP concentration. 
 When there are no tau molecules bound to the MT, motor proteins can only detach from the filament when they are in state 1. 
 So, as $\omega_h$ increases, the ATP concentration increases and the time that a motor is in state 2 becomes larger (if there are enough ATP molecules, the motor hydrolyzes one and changes to state 2 almost immediately). Thereby, the time duration of the motor runs in the MT increases. 
 On the other hand, in the presence of tau molecules, KIF1A motors can be forced to detach even if they are in state 2. 
 When a moving motor protein (either actively moving forward or diffusively moving towards either end) encounters a tau molecule  in its target site, it is forced to detach from MT with a probability $p_m$. 
 In this way, the presence of tau molecules shortens the averaged bound time. 
 The greater the values of tau concentration, $C_{tau}$, the larger the number of KIF1A-tau encounters and the stronger the effect of tau on the averaged bound time.
 This can be observed in Fig. \ref{fig:t}, where the values of averaged bound time are ordered with increasing value of $C_{tau}$ from top to bottom. 
 The results for $C_{tau}=0.005$, in which the tau concentration is very low, are similar to the case without tau, i.e., the averaged bound time increases with $\omega_h$. 
 For $C_{tau}=0.01$ and $C_{tau}=0.1$ the tau concentration is large enough to cancel the effect of the high ATP concentration and the averaged bound time remains almost constant for all $\omega_h$ values. 
 Moreover the averaged bound time for $C_{tau}=0.1$ presents a slightly decreasing tendency with increasing values of $\omega_h$ (averaged bound time decreases approximately from $ 0.22 s$ to $0.13 s$ )  which corresponds to a tau-ruled behavior. 
 A high ATP concentration accelerates the dynamic of the motors favoring the encounters with tau molecules and increasing the chances of tau mediated detachment.  

\begin{figure}[htb]
	\begin{center}
		\includegraphics[width=8.6cm]{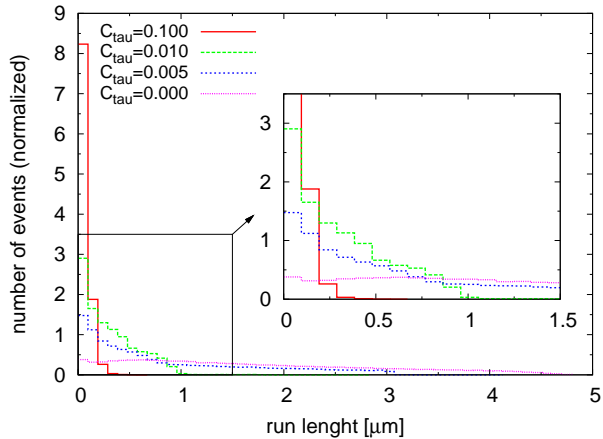}
		\caption{Histograms that show the normalized number of events for the run length of KIF1A motors. Model parameters: $L=600$, $p_a=0.67$, $p_m=0.5$, $\omega_d=\beta_1=\beta_2=0.0001[{ms}^{-1}]$, $\omega_s=0.145[{ms}^{-1}]$,$\omega_f=0.055[{ms}^{-1}]$,  $\omega_b=0.6[{ms}^{-1}]$,  $\gamma_1=\gamma_2=\delta=0$, $\omega_h = 0.2 [{ms}^{-1}]$ and $\omega_a=\alpha=0.00001 [{ms}^{-1}]$. Total number of MCSs = 2x$10^6$, after a thermalization of 2x$10^5$ MCSs. \label{fig:hist_l} }
	\end{center}
\end{figure}

 The effect of tau on KIF1A dynamics considering the averaged run length (calculated by averaging the distances that each motor travels between attachment and detachment) can be established in similar terms. 
 Firstly, we can note in Fig. \ref{fig:l} that in absence of tau molecules the averaged run length increases with $\omega_h$.  
 The ATP molecules are the fuel that KIF1A motors use to actively move toward the plus end of MT, so that a greater ATP concentration accelerates the dynamic of the motors enlarging the distance that the motors travel before detaching from MT. 
 In presence of tau molecules, the movement of KIF1A along the MT can give room to encounters with tau molecules so that a competition begins when we increase the ATP concentration. 
 On the one hand, as we already mentioned, the availability of ATP molecules favors longer motor run lengths, but, on the other hand, it also favors larger numbers of kinesin-tau encounters, which shortens  motor run lengths. 
 Therefore for $C_{tau}=0.005$ and $C_{tau}=0.01$ the run lengths are still  increasing functions of $\omega_h$ but for $C_{tau}=0.1$ it remains almost constant for the whole range of $\omega_h$. 
In this case, the tau concentration is large enough to compensate the higher ATP concentration with a greater number of motor detachments. 
 For the same value of $\omega_h$ (ATP concentration) the run length is always smaller for larger values of $C_{tau}$, that is,  the effect of tau is to shorten the distance that the motors travel along the MT.  
 Figure \ref{fig:hist_l} shows the differential effect of several tau concentrations on the KIF1A run length distributions. 
 Note that these results agree well with the experimental observations communicated by Dixit et al. in Ref. \cite{Dixit}.
 
\section{Conclusions}

 In this paper we have presented a novel stochastic model for the intracellular transport by KIF1A motor proteins that takes into account the effect of the MAP tau. 
 Our model is an extension of the model of Nishinari et al. for the trafficking of KIF1A without tau, and intend to preserve one of its virtues, that is, to relate the properties of the transport to experimentally controllable quantities. 
 In this way, we model the effect of one tau molecule by means of two parameters $p_a$ and $p_m$.
  The effect the tau molecule has in preventing kinesin motor binding to MT is characterized by $p_a$, and the effect it has forcing moving motor proteins to detach from the MT upon an encounter, by $p_m$. 
 These parameters values have been determined from experimental results obtained by Dixit et al. \cite{Dixit} and have proven to be robust from the point of view of the simulation. 
 That is, small changes on the values of these parameters do not modify the qualitative behavior of the observables discussed here.
 Moreover, changes in orders of magnitude on the parameters $p_a$ and $p_m$, change the value of the observables in the same order of magnitude (result not shown). 
 Our model coincides with the model by Nishinari et al. \cite{Nishinari} in the limit where no tau molecules are present in the filament, and the same is true for the analytical approximate high density results shown in Eqs. \ref{eq:r_limite} and \ref{eq:q_limite}. 
 The processivity of molecular motors, which is crucial for the intracellular transport, can be defined in several ways \cite{Parme_Jul_Prost}, two of that being the attachment lifetime of the motor proteins in the filament (bound time) and the mean length traveled by the motor proteins in a single run in the filament (run length). 
 We found significant effects on both  bound time and run length of KIF1A motors due to the presence of bound tau molecules in the MT. 
 The results shown here suggest that tau could play an important role in regulation of intracellular transport by kinesins.

\begin{acknowledgments}

	J.S. and P.W.L. want to thank SECyT--UNC (Argentina) for financial assistance. 

\end{acknowledgments}


\bibliography{Javier}{}


\end{document}